# Nonlocal Flat Optics

Kunal Shastri[1] and Francesco Monticone[1,*]

[1] School of Electrical and Computer Engineering, Cornell University, Ithaca, NY 14850, USA.

* e-mail: francesco.monticone@cornell.edu

**Abstract:** In electromagnetics and photonics, "nonlocality" refers to the phenomenon by which the response/output of a material or system at a certain point in space depends on the input field across an extended region of space. While nonlocal effects and the associated wavevector/momentum-dependence have often been neglected or seen as a nuisance in the context of metasurfaces, the emerging field of *nonlocal flat optics* seeks to exploit strong effective nonlocality to enrich and enhance their response. In this Review, we summarize the latest advances in this field, focusing on its fundamental principles and various applications, from optical computing to space compression. The convergence of local and nonlocal flat optics may open exciting opportunities in the quest to control light, in real and momentum space, using ultra-thin platforms.

In the first century AD, Seneca, the Roman philosopher, noted that "*Letters, however small and indistinct, are seen enlarged and more clearly through a globe or glass filled with water*" [1]. For hundreds of years, despite sophisticated technological improvements, lenses – the quintessential optical component – were not conceptually very different from Seneca's globe from two millennia ago, as they both were based on the basic mechanism of light refraction by a curved dielectric object. From a wave optics viewpoint, the emerging wavefront is shaped by a continuous, *transverse* profile of phase delay acquired through propagation in the lens material. In recent centuries, diffractive dielectric optics has then provided a strategy to "flatten" lenses by introducing a phase profile modulo $2\pi$, albeit still based on light propagation inside a dielectric material, and at the expense of typically higher aberrations. More recently, over the past few decades, thanks to tremendous advances in nanotechnology, it has become possible to impart a specific phase profile not (or not only) through propagation, but via the local scattering response of engineered nanostructures ("nano-antennas" or "meta-atoms") in a planar array [2]-[11]. These *metasurfaces* – the two-dimensional version of metamaterials – provide greater design flexibility than standard diffractive lenses, as their nanostructures can be designed with a variety of shapes and materials to create a tailored optical response in terms of amplitude, phase, polarization, frequency, etc. [11] Modern diffractive elements and metasurfaces are the subject of the field of "flat optics", which is providing a route towards the miniaturization of various optical systems for applications spanning from imaging and augmented reality, to sensing and endoscopy.

Despite the ongoing success of flat optics and the ever-increasing interest in this field from both academia and industry, the range of functionalities that can be achieved with flat optic devices is not fundamentally different from those available in conventional optical systems, for example, beam deflection, focusing, wavefront and polarization transformations. This is because

metasurfaces – akin to refractive lenses, transmissive/reflective masks, and spatial light modulators – are *local* devices, namely, they control wave transmission/reflection as a function of position, locally and pointwise (approximately), based on an engineered, transversely inhomogeneous structure. In other words, the output of these individual optical elements at some position on their output plane only depends on the input at a specific point (as illustrated in Fig. 1a), at least approximately. In this sense, all these devices are not fundamentally different from Seneca's transversely inhomogeneous globe lens. A wide range of functionalities that go beyond position-dependent wavefront transformations are inaccessible by individual local devices. For instance, a single refractive optical element or standard metasurface cannot perform, by itself, image processing operations, such as edge/feature detection, that depend on an *extended* portion of the spatial profile of the incident beam and therefore on its plane-wave spectrum. While this type of operations can be realized with Fourier optics, systems of this kind require multiple elements, such as lenses, phase/amplitude masks, and large free-space volumes, creating a system several hundreds or thousands of times longer than the wavelength, even if all the lenses were replaced by metalenses. Indeed, the miniaturization and planarization of lenses is not sufficient to miniaturize, in terms of thickness, the vast majority of free-space optical systems, which usually comprise multiple elements separated by empty volumes. More broadly, we note that most optical systems (e.g., a camera) are *globally* nonlocal in their input-output relation, namely, their output at a certain point (e.g., a pixel) needs to depend on the input in an extended spatial region (e.g., the aperture). It is clear, therefore, that such a globally nonlocal functionality cannot be implemented with a single local device. To address these challenges, increasing attention has been recently devoted to *nonlocal* metasurfaces, which are not limited to a local position-dependent response. Instead, their response/output at a certain point on the output plane depends on the input field across a region of space, as illustrated in Fig. 1b.

As further discussed in Box 1, in the context of the macroscopic electrodynamics of continuous media, a nonlocal response occurs naturally in many materials, in the form of a nonlocal input-output relation between the induced polarization density and the applied field, which physically originates from polarization or charge-density waves that propagate across the material, creating a polarization response in regions potentially quite far from where the incident field is applied. This effect is common, albeit weak, in natural materials such as conductors [12], and it can be significantly enhanced in artificial metamaterials, even if all constituent materials are local, through the engineered interaction between neighboring elements and/or by using meta-atoms with a multipolar or bi-anisotropic response [14],[15]. Similarly, in metasurfaces, formed by planar arrays of polarizable meta-atoms, an effective nonlocal response is always present, but it is usually weak and, until recently, was usually treated as an issue to mitigate, because a metasurface with non-negligible nonlocality would respond slightly differently depending on the specific spatial profile of the incident field. Conversely, the field of nonlocal flat optics seeks to boost and leverage this property, and the new degrees of freedom it affords, to enable new functionalities and overcome some of the limitations of conventional optics and local flat optics.

The potential of metasurfaces with an effective, nonlocal, input-output response can be better appreciated by moving our viewpoint from real space to momentum/wavevector space, namely, spatial Fourier space, corresponding to a plane-wave expansion of the fields. As illustrated in Fig.

1b,d, if the response of a material or structure is nonlocal in real space (spatial impulse response is not a delta function), the corresponding transfer function in momentum space is not constant. In other words, the response of the material or structure is momentum-dependent, i.e., it depends on the wavevector **k** of the applied field and, therefore, on its angle of incidence $\theta$. This property is also commonly referred to as "spatial dispersion", in analogy to temporal (frequency) dispersion [12],[13]. As discussed in the following, many applications can benefit from the new degrees of freedom afforded by nonlocality – angle-dependent amplitude and phase, transverse energy channeling, non-trivial frequency-momentum correlations, etc. – especially if implemented in planar metasurface form.

**Physical mechanisms and platforms for nonlocal flat optics**

Nonlocal metasurfaces require very different design principles and methodologies compared to local devices. While standard metasurfaces, as well as most optical components, are based on transversely varying distributions of some geometrical or material parameters, purely nonlocal metasurfaces do not necessarily require a position-dependent response and, therefore, they can be transversely homogenous or periodic, whereas they are typically inhomogeneous in the longitudinal direction.

What *physical mechanisms*, materials, and structures are then needed to implement a strongly nonlocal momentum-dependent response in metasurface form? Since natural materials only exhibit weak nonlocality (Box 1), one needs to rely on artificial effective nonlocality based on specific structures and arrays of elements. All nonlocal metasurfaces discussed in this review are themselves formed by local materials, and thereby in the microscopic sense the electrodynamics is local and momentum-independent. We also note that a weakly momentum-dependent response, per se, is easy to implement and is actually ubiquitous in the context of reflection and transmission from generic planar structures. For a single interface between two materials, for instance, a momentum/angle-dependence originates from the scaling with angle of the transverse wave impedance of incident, reflected, and transmitted plane waves. Specifically, the transverse wave impedance scales as $\cos\theta$ for transverse-magnetic waves ("transverse" with respect to the incidence plane) and as $1/\cos\theta$ for transverse-electric waves [16]. This results in an angle-dependent response, particularly at grazing angles, as in the case of Fresnel reflection/transmission at a planar interface. This form of momentum-dependence is however too weak to be of much practical use and is often considered an unwanted feature for metasurfaces and antenna arrays [17].

If an effective nonlocal response is desired, a possible strategy is to use metasurfaces formed by multipolar meta-atoms, as their multipole moments, and the resulting scattered fields, depend on the spatial derivatives of the applied field; however, for small optical meta-atoms their multipolar polarizability is typically weak. A stronger nonlocal response may be obtained with meta-atoms that cross many unit cells of the metamaterial or metasurface to effectively connect the response of cells that are far apart in the structure, as done in [18],[19],[20]. Along similar lines, even stronger artificial nonlocal effects require the interaction of the incident field with fully delocalized modes or, in other words, the creation of a wave of induced polarization traveling across the metasurface over relatively large distances, which translates, through a Fourier transform, in a particularly strong, tailorable dependence on the incident wavevector/momentum. In the recent literature, two related strategies have been employed to realize these effects in a planar thin platform: (i) multiple reflections between thin homogenous or homogenized layers in multilayer structures (Fig. 2a), or (ii) excitation of guided modes in arrays, metasurfaces, or photonic crystal slabs (Fig. 2b,c). The first strategy is based on the intuitive fact that multiple reflections between interfaces create a much stronger angular (and frequency) dependence than in the case of a single interface thanks to interference effects between reflected/transmitted waves at different interfaces. Optimizing the thickness and material of the layers, in both periodic and non-periodic stacks, allows tailoring the angular/frequency dependence and the corresponding spatial impulse response in real space. Clearly, in order to make this nonlocal response stronger, a large number of light "bounces" within the structure is beneficial, which can be understood both in terms of a stronger angular dependence due to more complicated interference effects and in terms of a larger transverse wave propagation within the structure at oblique incidence, as illustrated in Fig. 2a (such that fields over a wider region of the input plane can contribute to the output at a given point). A structure supporting multiple round-trips for light is effectively a resonator, e.g., a Fabry-Perot resonator or a Gires–Tournois etalon. The number of round-trips, and therefore the stored energy in the structure, is strongly angle- (and frequency-) dependent around each resonance, providing a strategy to synthesize a strong nonlocal response by optimizing multiple resonances within the structure as well as their coupling. While off-resonant multilayer structures also offer moderate nonlocal effects, mostly determined by the different penetration depth ("turning point" [21]) for light incident at different angles, strongest spatial dispersion is certainly obtained through multiple internal round-trips and the resulting angle-dependent stored energy [22]. These observations also already point at one of the main challenges, and opportunities, of strongly nonlocal structures of this type, where spatial and frequency dispersion are not independent, as stored energy strongly varies both with frequency and angle. It is, therefore, usually challenging to create broadband nonlocal devices [23], whereas high spectral and angular selectivity may be useful to enable multiple functionalities with a single metasurface or non-trivial wavevector-frequency correlations, as discussed in the next sections. In fact, these concepts were employed, even before the advent of metasurfaces and flat optics, to design multilayer thin-film structures in which strong spatial and frequency dispersion were used to create "super-prism" effects, separating beams of different wavelengths by a large transverse spatial shift [22],[24]. More recently, strongly nonlocal multilayer structures have been used to realize spatial-frequency filters for analog optical processing [25] and angle-dependent phase shifters for space compression [23],[26],[27], as further discussed in the following.

Each of the resonances of a multilayer structure is associated with a guided mode of the structure propagating in the transverse direction (parallel to the interfaces) and attenuated by radiation leakage, i.e., a leaky mode [28]. If radiation and absorption losses are not too high, a resonant frequency of the structure is close to the real part of the eigenfrequency of the corresponding eigenmode. This fact clearly underlines the connection between the first and second strategies, mentioned above, to create strong artificial nonlocality. Any structure supporting one or multiple guided leaky modes is naturally nonlocal, as the response of the structure, e.g., its transmission, will exhibit guided-mode resonances following the momentum-frequency dispersion of these modes, $\omega(\mathbf{k}_t)$, where $\mathbf{k}_t$ is the transverse wavevector/momentum. Physically, these resonances occur when transverse-momentum conservation ("phase-matching condition") is satisfied between the incident wave and the guided mode (one of its space harmonics if the structure is periodic [28]), and they typically manifest as asymmetric, sharp spectral features known as "Fano resonances" [29], which stem from the interference between the slowly varying off-resonant response of the structure (the "direct" transmission through the structure) and the Lorentzian guided-mode response providing an "indirect" transmission channel (Fig. 2b,c). For a single resonance, one can write: $T(\omega, \mathbf{k}_t) = t_d(\omega, \mathbf{k}_t) + f\gamma(\mathbf{k}_t)/(i(\omega - \omega_0(\mathbf{k}_t)) + \gamma(\mathbf{k}_t))$, where $t_d$ is the direct transmission coefficient, $f$ is the complex amplitude of the resonant mode, and $\omega_0$ and $\gamma$ are the center frequency and width of the guided-mode resonance, respectively [30]. While both the direct and indirect transmission are frequency- and wavevector-dependent, this dependence is much stronger for the guided-mode-mediated channel. By operating close to a guided-mode resonance, it is therefore possible to create a strongly nonlocal angle-dependent response dictated by the interaction between incident wave and guided mode at a certain frequency and angle of incidence. In different terms, this strategy leverages the amplitude and phase contrast one can obtain around a resonance as a function not of frequency, as usually done to create filters and resonance-based phase shifters, but of transverse momentum. Thus, engineering the band diagram and the linewidth of the guided modes of a planar structure, as well as its symmetries as discussed below, provides a direct approach to implement a large variety of nonlocal functionalities. As illustrated in Fig. 2, platforms of choice in this context include multilayer structures, metasurfaces, gratings, and photonic crystal slabs, where the two-dimensional band-diagram of guided modes can be engineered by playing, respectively, with the sequence of layers and their material composition, the shape and orientation of subwavelength meta-atoms, the periodicity of the structure, or a combination of these strategies.

**Passivity, reciprocity, and mirror symmetries**

Engineering the response of a device in momentum space is significantly more challenging than in real space. For example, while it is trivial to realize non-symmetric responses as a function of transverse position $x$, it is difficult to do so with respect to transverse wavevector. More generally, only certain nonlocal transfer functions may be permitted by the fundamental symmetries of the considered system and other physical constraints. Consider, for example, a nonlocal metasurface operated near a transmission zero. A Taylor series expansion of $T(k_x)$ around this zero, located for example at $k_x = 0$ (normal incidence), reads $T(k_x) \approx c_1 k_x + c_2 k_x^2 + c_3 k_x^3 + \ldots$, where we assumed $k_y = 0$ for simplicity. The nonlocal metasurface design problem can then be interpreted as the problem of engineering a structure with the required expansion coefficients to match a transfer function of choice. General principles and symmetries impose strict conditions on what transfer functions can be implemented: (i) Energy conservation in a passive system implies that $|T(k_x)| \leq 1$, so any desired transfer function should be scaled to meet this condition. (ii) The transmission phase $\angle T(k_x)$ is limited within the range $[-\pi/2, \pi/2]$ (not the entire $2\pi$ phase range) if a single resonance is employed, as seen for instance in the expression for the guided-mode-resonance transmission in the previous section; if a larger phase response is desired, multiple resonances need to be used. (iii) Reciprocity (equivalent to time-reversal symmetry in the lossless case) implies that the dispersion relation of the guided modes is symmetric for opposite transverse wavevectors, i.e.,

$\omega(k_x) = \omega(-k_x)$. Since the transmission response is dominated by the presence of guided-mode resonances, this dispersion symmetry may suggest that the transmission coefficient $T(k_x)$ should be the same for opposite values of $k_x$. However, the behavior of $T(k_x)$ is more subtle and depends on both reciprocity and the mirror symmetries of the structure, as nicely discussed in Ref. [17]. Consider the illustrations in Fig. 2d-g, where different examples of meta-atoms of a nonlocal metasurface are illustrated together with four possible directions of incidence and transmission. If the structure has mirror symmetry with respect to the vertical z-axis (i.e., with the z-axis as the mirror plane), as in Fig. 2d, it is then clear that transmission from direction 1 to 4 is the same as transmission from 2 to 3, $T_{1 \to 4} = T_{2 \to 3}$, as they are related by a mirror operation; therefore, using the simpler notation previously used for transmission of light from a specific side, $T(k_x) = T(-k_x)$. Interestingly, we can reach the same conclusion even if the vertical-mirror symmetry is broken, but the structure still has horizontal-mirror symmetry (i.e., with the x-axis as the mirror plane) and is reciprocal [17]. In this case (Fig. 2e), $T_{4 \to 1} = T_{2 \to 3}$ since they are related by a mirror operation. By combining this equality with the reciprocity condition $T_{1 \to 4} = T_{4 \to 1}$ (detector response is the same when the source and detector positions are switched [12]), we can conclude that $T_{1 \to 4} = T_{2 \to 3}$ and, therefore, $T(k_x) = T(-k_x)$, as before. Thus, an asymmetric transmission response may be achieved only by breaking reciprocity or the horizontal-mirror symmetry in addition to the vertical one. Indeed, for a reciprocal meta-atom lacking both horizontal- and vertical-mirror symmetry, as in Fig. 2f,g, reciprocity only implies that $T_{1 \to 4} = T_{4 \to 1}$ and $T_{2 \to 3} = T_{3 \to 2}$. Instead, reciprocity does *not*, in general, imply that $T_{1 \to 4} = T_{2 \to 3}$, which therefore makes an asymmetric response, $T(k_x) \neq T(-k_x)$, possible. We also note that, interestingly, this is not possible for a nonlocal metasurface operating in reflection, as reciprocity strictly implies the equality of reflection upon exchange of source and detectors, $R_{1 \to 2} = R_{2 \to 1}$ and $R_{3 \to 4} = R_{4 \to 3}$; therefore, $R(k_x) = R(-k_x)$ [31].

To summarize, any reciprocal array of elements with either horizontal- or vertical-mirror symmetry, or both, exhibits the same transmission response for opposite transverse wavevectors. Hence, only the even terms in the series expansion of $T(k_x)$ are non-zero, and the first term around a transmission zero is always the quadratic one, i.e., $T(k_x) \approx c_2 k_x^2 + \ldots$. As a result, since the momentum-dependent response can be approximated as quadratic sufficiently close to *any* transmission zero of *any* planar structure with these symmetries, it is not surprising that this is the most commonly studied nonlocal transfer functions in the literature [17],[25],[32]-[36]. The practical interest in this response stems from the fact that the second-order derivative operator $\alpha\, d^2/dx^2$ in real space corresponds to the transfer function $T(k_x) = -\alpha k_x^2$ in momentum space. Hence, this important operator, which can be used to perform image differentiation and analog edge detection on an incident wavefront, can be readily implemented by operating near a transmission (or reflection) zero of the structure and optimizing it to minimize higher-order terms. Another application of a quadratic nonlocal response, but for the transmission phase instead of the amplitude, is to emulate the angle-dependent phase response of a free-space volume, creating so-called nonlocal "spaceplates" for space compression [23],[26],[27],[37]-[39], as further discussed in the next section. In contrast with these symmetric responses, an different transmission for opposite wavevectors requires breaking *both* horizontal- and vertical-mirror symmetries (or reciprocity), in which case the odd terms in the series expansion are non-zero, and the first term around a transmission zero is the linear one, i.e., $T(k_x) \approx c_1 k_x + \ldots$. This may also be interpreted as the combination of an asymmetric *direct* transmission $t_d$ through a structure lacking horizontal and vertical symmetries, and the *indirect* transmission through a guided leaky mode, which is inherently symmetric in reciprocal structures, as it follows $\omega(k_x) = \omega(-k_x)$. Such a nonlocal response can be used to realize, for instance, the first-order spatial derivative operator, $\alpha \, d/dx$, corresponding to an odd-symmetric transfer function $T(k_x) = i\alpha k_x$ [17],[33].

**Applications of nonlocal flat optics**

The new degrees of freedom afforded by nonlocal metasurfaces, constrained by the symmetries and physical principles discussed above, provide new opportunities for the field of flat optics. In this section, we briefly review the main applications of these concepts in different areas.

Analog Image Processing and Optical Computing. The ability of nonlocal metasurfaces to control and shape the wavevector-dependent transmission/reflection amplitude makes them ideal to realize ultra-compact spatial frequency filters and perform spatial operations on the incident wavefront, which is particularly relevant for image processing tasks. For instance, a nonlocal metasurface designed as a high-pass spatial frequency filter will only transmit the components of the incident image with high spatial frequency (large wavevector), corresponding to small, sharp spatial features of the image, e.g., edges and corners, while filtering out smoother details (Fig 1e, and Fig. 3a,b,c). The already mentioned transfer functions, $T(k_x) = -\alpha k_x^2$, and $T(k_x) = i\alpha k_x$, which can be used to perform spatial differentiation, are indeed examples of high-pass spatial frequency filters that have been demonstrated for analog edge enhancement/detection, both in classical electromagnetism/optics [17],[25],[32]-[36] and quantum optics [40]. Other important operations that nonlocal metasurfaces of this type may enable include low-pass spatial filtering to smooth out images, or band-pass and band-stop filters to enhance or suppress artifacts and image features with a certain spatial periodicity [41]. In this way, various image processing tasks may be performed analogically on the physical layer, at the speed of light, and with negligible power consumption if the employed materials have low optical losses. We also speculate that these ideas may also be very relevant in conjunction with software approaches for feature extraction.

More broadly, these concepts are not limited to image processing, but can be applied to any computational task on signals with information encoded in the spatial domain, and may be extended to include the temporal domain by engineering the frequency dispersion of the structure. Indeed, nonlocal metasurfaces may perform, in principle, all the processing operations of Fourier optics, but in the form of a single, compact, optical element [25],[42]. While Fourier optical systems involve lenses to perform a Fourier transform and a phase/amplitude mask to apply an operation of choice [43], wave-based computation based on nonlocal metasurfaces removes the requirement of an explicit transformation to momentum space, as the wavevector-dependent transfer function, i.e., the Fourier transform of the spatial impulse response (Green's function) of the system, is directly engineered to match the Fourier transform of the desired spatial operator [25],[42]. Nonlocal flat optics may therefore enable ultra-compact, planar, analog optical computing platforms, in which mathematical operations are performed passively, "on the fly," as the wave propagates through the structure.

Enhancing the performance of metasurfaces and metalenses. The practical and fundamental limitations of local metasurfaces have been well studied over the past decade. An important challenge that this body of work has identified is the fact that, even at a single frequency, passive lossless metasurfaces based on the generalized laws of reflection/refraction [4] cannot create arbitrary wavefront transformations with unitary efficiency, particularly in reflection and when extreme transformations are desired, such as beam steering at large angles [44],[45],[46]. The issue ultimately originates from the requirement that power conservation is satisfied at the local level, i.e., at each point on the metasurface power is not absorbed or generated, but simply reflected or transmitted. This requirement may be relaxed in nonlocal metasurfaces, in which energy can be channeled transversely along the metasurface through the excitation of a guided leaky mode and then reradiated over an extended region (Fig. 3d) (incidentally, we note that this energy channeling effect is analogous to the enhanced Goos–Hänchen shift in structures supporting guided leaky modes [28]). In this case, energy appears to be lost or gained at the local level, even if the overall structure is perfectly lossless and satisfies global power conservation. Hence, thanks to the design flexibility afforded by strong nonlocality, which allows optimally utilizing the available energy across the surface, perfect control of anomalous reflection and refraction was demonstrated in metasurfaces that emulate the required nonlocal response via the interference with auxiliary evanescent modes [46] or leaky modes [45],[47]. Similar ideas have also been applied to different domains of wave physics, creating nonlocal metasurfaces for efficient manipulation of acoustic [48] and elastic [49] waves. In addition to efficient wavefront transformations, design strategies based on, or inspired by, nonlocal effects have been employed to create a large variety of enhanced flat-optic devices and new functionalities, including thin absorbers and anti-reflection coatings with improved angular performance [50]-[52], metalenses free of chromatic and angular aberrations [53], and parity-time symmetric nonlocal metasurfaces for all-angle coherent perfect absorption, negative refraction, and volumetric imaging [54]-[56].

Multi-functional metasurfaces with high angular and spectral control. The ability to realize optical functions with selective angular and spectral control is highly desirable for various applications, for example in the context of transparent displays and augmented reality systems, which need to offer an unperturbed view of the world while displaying information and, perhaps, imaging the eye (eye-tracking) at the same time. Although conventional local metasurfaces can be designed to have specialized transfer functions at discrete angles and frequencies, for example, using multimode resonators as meta-atoms [57], their response is only weakly angle-dependent, and relatively broadband since the quality factor of small optical scatterers is relatively low. Nonlocal metasurfaces and photonic-crystal slabs offer an ideal solution to this problem thanks to their strong spatial and spectral dispersion. Several recent works have employed these platforms to selectively manipulate light at different angles and wavelengths following the sharp dispersion of a guided mode [58]-[65]. In this context, the concept of quasi-bound states in the continuum (BICs), which are an extreme form of guided leaky mode, almost decoupled from free-space radiation by symmetry or interference effects, has proven particularly fruitful to realize structures with very large Q factors and angular selectivity [66]-[69]. By combining nonlocal design strategies for spectral and angular control with spatially varying transverse patterns for local control of the wavefront, it is then possible to create multifunctional metasurfaces producing spatially shaped wavefronts at different selected angles and/or wavelengths. For example, a metasurface may be designed to focus specific colors in reflection while transmitting other wavelengths, as illustrated in Fig. 1e, which is an important operation for transparent displays. A systematic method to independently control the local and nonlocal interactions in a metasurface supporting quasi-BICs, based on analyzing symmetry-preserving perturbations to the nonlocal modes, was recently proposed in [59]-[60] and used to implement wavefront-selective metasurfaces [63] and spectrally selective metalenses for transparent displays [65]. Finally, while for nonlocal metasurfaces relying on guided-mode resonances the response at different wavelengths is not fully independent as it follows the band structure of the relevant mode, it was shown in Ref. [64] that the introduction of frequency-dependent material absorption in a guided-mode-based structure can be used to effectively decouple the optical functions at different wavelengths. Such a nonlocal metasurface with spectrally decoupled response was used to create eye-tracking glasses that provide an unperturbed view of the world in the visible spectrum, while deflecting near-infrared light to allow imaging and tracking of the eye (Fig. 3e).

Thermal metasurfaces. Light emitted by objects as thermal radiation, including solar radiation, is characteristically incoherent, both temporally and spatially, resulting in broadband and omnidirectional light emission [70]. However, several works over the past two decades have demonstrated that partially coherent thermal radiation can be generated from structured surfaces supporting extended guided leaky modes [28], namely, structures with a strong nonlocal response. Typical platforms that have been employed in this context are arrays of coupled resonators [71],[72], layered media [73],[74], and photonic crystal slabs [75]-[76]. More recently, full control over thermally emitted light was theoretically demonstrated in Ref. [77] based on metasurfaces that combine nonlocal interactions, via a thermally populated guided mode, and local interactions, determined by subwavelength structuring. In this way, by locally tailoring the radiation leakage from the extended guided mode, a partially coherent wavefront of choice can be obtained (Fig. 3f).

Creating optical wave packets with non-trivial space-time coupling. Light bullets are dispersion-diffraction-free optical wave packets that propagate in free space without deformation and with a group velocity $v_g$ that may differ from the free-space speed of light $c_o$. This effect is obtained by shaping the wave packet such that its temporal and spatial frequency components are strictly related as $\omega = v_g k_z + b$, where $b$ is a constant. Previous efforts to realize such a "space-time coupling" were based on the use of spatial light modulators and ultra-fast pulse shaping techniques [78]-[79]. Interestingly, nonlocal flat optics provides an ideal solution to create light bullets within a much more compact platform: The frequency- and wavevector-dependent transfer function of a nonlocal structure, e.g., a photonic-crystal slab, supporting a suitable guided-mode resonance can be utilized to filter an incident pulse in frequency-wavevector domain and impart the required space-time correlation (Fig. 3g). Thus, the incident pulse is directly converted into a light bullet with properties controlled by the dispersion, linewidth, and polarization texture of the guided mode.

Space Compression with Nonlocal Spaceplates. Flat optical components such as local metalenses can substitute and miniaturize conventional refractive optical elements in addition to providing greater design flexibility. However, as mentioned in the Introduction, this leads to significant, but ultimately incremental improvements in terms of size and weight for complex optical systems. Even more drastic reductions in the form factor of free-space optical systems (e.g., a cellphone camera or a telescope), and potentially the complete monolithic integration of all optical elements, are only possible if the free-space volumes between consecutive optical elements and solid-state detectors can also be reduced, or completely removed, without changing the optical performance and functionality of the system. Conceptually, this requires emulating the angle- and distant-dependent phase gained by a plane wave propagating a distance $L_{eff}$ in free space, $\phi(k_t) = L_{eff}\sqrt{\omega^2/c_0^2 - k_t^2}$, using a device with a smaller thickness $L < L_{eff}$, thereby effectively "compressing space" for light propagation by a factor of $L_{eff}/L$. Since the required phase response is inherently angle-dependent, nonlocal metasurfaces are well-suited to implement this optical function within a thin planar structure, realizing space-compression devices known as "spaceplates" [23],[26],[27],[37]-[39].

The same platforms described above for spatial-frequency filtering, namely, photonic crystal slabs and multilayer structures, can also be used to realize spaceplates (Fig. 4). However, instead of engineering these structures to control the wavevector-dependent transmission *amplitude,* the goal here is to control the wavevector-dependent transmission *phase* to match the response of a longer volume of free space, while keeping the amplitude as high as possible. If only a relatively narrow angular range around normal incidence is of interest, space compression can be realized by designing the nonlocal metasurface such that the series expansion of the required phase response, $\phi(k_t) \approx k_0 L_{eff} - (L_{eff}/2k_0)k_t^2$, is matched by the series expansion of the metasurface transmission phase, $\angle T(k_t) \approx \phi_0 + c_2 k_t^2$, with a quadratic coefficient as large as possible to realize a large compression factor [37]. The nonlocal response for space compression can also be interpreted as the realization of a transverse shift for incident waves, as shown in Fig. 4a,d, which allows, for example, focusing light at a shorter distance without changing the focal length of the system, which is an intrinsic property of the local optical elements. Remarkably, this leads to the complete decoupling of the focal length, and therefore the obtained image magnification and other imaging properties, from the actual length of the imaging system (this can also be obtained with a standard telephoto lens, but only partially [26]). Interestingly, as noted in Ref. [23], this space-compression mechanism is the dual process of a lens: whereas a lens is a local, transverse-position-dependent device that changes the angle of a light beam, a spaceplate is a nonlocal, angle-dependent device that changes the transverse spatial position of a light beam.

Different spaceplate implementations reported in the literature have resulted in different, and somewhat complementary, performance. Spaceplates based on photonic-crystal slabs [37], leveraging the sharp dispersion of a Fano resonance, can realize very large compression ratios, but are limited with respect to bandwidth and angular range (numerical aperture). Spaceplates based on multilayer thin films [26],[27],[38],[39] , on the other hand, typically have more modest compression ratios, but larger spectral and angular ranges. These tradeoffs between compression ratio, bandwidth, and numerical aperture are, in fact, fundamental for any spaceplate device, as they ultimately originate from the number of resonances/modes within the structure that can be used to create the required nonlocal response, as demonstrated in Refs. [23],[27],[39]. However, current spaceplate designs are still relatively far from their fundamental performance bounds [23], leaving large room for improvement using more sophisticated design strategies, including topology optimization and inverse design approaches [38],[53]. Future high-performance spaceplates may play a crucial role in the quest for the miniaturization of free-space optical systems for focusing, imaging, sensing, and various other applications.

**Conclusion and outlook**

While the field of nonlocal flat optics is still in its infancy, it has the potential for transformative advances in the development of ultra-compact planarized optical systems, enriched by new functionalities difficult or impossible to realize with local metasurfaces. Recent works in this area have highlighted the many opportunities that the new degrees of freedom afforded by strong nonlocality may enable, as described in this Review. For this potential to be fully realized, however, more work is needed to address some of the remaining challenges of these novel flat-optic devices. For instance, the operating bandwidth of these platforms is usually narrow since frequency dispersion and spatial dispersion are generally related, as strong nonlocality typically involves resonant effects. While for some applications spectral selectivity is a desirable feature, as discussed in the previous section, bandwidth may be an issue for nonlocal devices for imaging applications. Broadening the bandwidth, without a negative impact on the transmission/reflection efficiency, should be one of the main goals of future work in this emerging field. More broadly, a better understanding of the fundamental limits and tradeoffs between thickness, efficiency, bandwidth, numerical aperture, and other parameters and metrics is crucially important for further development in this area. More experimental work is also needed, especially in some of the most recent areas of this field, e.g., space compression, to better assess the practical feasibility of these ideas. For instance, the experimental demonstration of a broadband nonlocal spaceplate at optical wavelengths would be an important milestone in this field. Fortunately, the structural simplicity of nonlocal metasurfaces, typically based on stacks of dielectric films, arrays of meta-atoms, and photonic-crystal slabs, and the fact that exotic material properties are not usually needed, make them suitable for fabrication with standard techniques, as already demonstrated in a few recent papers, e.g., [33],[35],[61],[64],[65]. Finally, we note that an important missing element in current research on nonlocal metasurfaces is a systematic investigation of how to make them tunable and reconfigurable. Compared to local flat optics, for which simple pointwise control of the optical properties directly maps into a change in their position-dependent response, reconfigurability in momentum space is more complex, as every individual element may contribute to the overall response of the nonlocal structure. New design approaches and actuation strategies may therefore be needed to endow nonlocal flat optics with reconfigurability.

If these remaining challenges are resolved, the combination of local and nonlocal components may represent the future of the field of flat optics. The synergy of position-dependent and momentum-dependent control may enable arbitrary and lossless spatial transformations of electromagnetic fields [81] in a monolithic flat-optic platform. In this way, as some of the works reviewed in this article already suggest, nonlocality-enhanced metasurfaces, or hybrid local-nonlocal metasurfaces, potentially designed using advanced inverse-design techniques, may (i) go beyond some of the limitations of local metasurfaces and conventional optical designs for wavefront transformation, wide-field-of-view imaging, selective angular and spectral control, etc.; (ii) enable flat optic systems with completely new functionalities, such as feature detection, thermal radiation shaping, and the creation of space-time light bullets; and (iii) drastically reduce the thickness of an entire optical system (lenses, waveplates, free space between them, etc.), enabling the realization of, for instance, paper-thin monolithic long-focus cameras, where the entire optical layer is implemented in the form a single, fully solid-state, thin device integrated directly on top of an electronic sensor. Looking further ahead, we also speculate that the convergence of these ideas with modern inverse-design methods, computational imaging, and deep learning may also open vast opportunities, as it is already happening for local flat optics [82]-[85].

To conclude, if its promise is fully realized, nonlocal flat optics may represent an important step forward in the thousand-years-old evolution of optics, expanding the flat-optic design toolbox with new approaches, and opening opportunities for transformative advances in the way we control light, both in real space, momentum space, and frequency domain, for countless applications.

**Box 1: Natural nonlocality in the macroscopic electrodynamics of continuous media**

Natural nonlocal effects, while usually weak, are common in macroscopic electrodynamics. In a material, the induced polarization density $\mathbf{P}(t,\mathbf{r})$ and displacement field $\mathbf{D}(t,\mathbf{r})$ depend on the applied electric field $\mathbf{E}(t',\mathbf{r}')$ at previous times $t' \leq t$ (temporal/frequency dispersion) and in some spatial region around point $\mathbf{r}$ (spatial dispersion, nonlocality). For linear time-invariant materials, this nonlocal (in time and space) input-output relation can be expressed as [12]

$$\mathbf{D}(t,\mathbf{r}) = \varepsilon_0 \mathbf{E}(t,\mathbf{r}) + \int_0^\infty \int \varepsilon_0 \tilde{\chi}_e(\tau;\mathbf{r},\mathbf{r}') \mathbf{E}(t-\tau;\mathbf{r}') dV' d\tau \ , \tag{1}$$

where $\varepsilon_0$ is the permittivity of free space and $\tilde{\chi}_e$ is the time-domain electric susceptibility of the material, which is generally a tensor. If the material is macroscopically homogenous, the spatial integral becomes a standard convolution integral. Then, by Fourier transforming (1) in time and space, corresponding to an expansion of the fields in monochromatic plane waves, one obtains

$$\mathbf{D}(\omega,\mathbf{k}) = \varepsilon_0 \left(1 + \chi_e(\omega,\mathbf{k})\right) \mathbf{E}(\omega,\mathbf{k}) \ , \tag{2}$$

where $\chi_e$ is the Fourier transform of $\tilde{\chi}_e$, $\omega$ is the angular frequency, and $\mathbf{k}$ the field wavevector. According to Eq. (2), natural spatial dispersion can be described as a dependence of the material properties on the field wavevector/momentum. The length scale of nonlocality, i.e., the distance $|\mathbf{r}-\mathbf{r}'|$ at which a polarization response is present, is inversely proportional to the width of the function $\chi_e$ in momentum space (Fourier uncertainty principle). While in ordinary dielectrics spatial dispersion is very weak, as this distance is small, in conductors (e.g., plasmas and metals), the kernel of the integral operator in (1) may extend to distances much greater than atomic dimensions owing to the movement of free carriers according to their mean velocity, mean free-path, and diffusion coefficient [12]. Nonlocal effects are particularly important for conductors in the presence of strong currents of drifting electrons, which induce a Doppler shift in the frequency dispersion of the permittivity proportional to $\mathbf{k}\cdot\mathbf{v}$, where $\mathbf{v}$ is the electron drift velocity [86]-[88]. The study of nonlocal effects is indeed one of the most important areas of research in the field of plasmonics. Since electromagnetic fields can be highly localized and confined in plasmonic structures, and hence they contain large-wavevector components, even weak, natural, nonlocal effects become important, as predicted by Eqs. (1) and (2). Different models of nonlocality for conductors have been developed based on phenomenological theories [89], hydrodynamic models with or without diffusion [90], and Landau damping [91]. Interestingly, it has been shown that nonlocal effects ultimately limit the level of field confinement and enhancement in plasmonics [92] and are important to determine under what conditions unidirectional modes exist in nonreciprocal plasmonic systems [93],[94]. For additional details on spatial dispersion in natural media and three-dimensional metamaterials, including the appearance of additional modes, the need for additional boundary conditions at interfaces, and the implications of nonlocality for the active and nonlinear optical properties of (meta)materials, we refer the readers to the relevant literature [12]-[15],[89]-[91],[95]-[97].

Nonlocality is even more fundamentally woven in the fabric of macroscopic electrodynamics than the previous discussion may suggest. Indeed, since Faraday's law relates the magnetic field to the spatial derivatives of the electric field, any effect produced by an applied magnetic field can be equivalently treated, at the level of classical macroscopic electrodynamics, as a manifestation of nonlocality [12]. Even more generally, the mapping from the microscopic Maxwell's equations in an arbitrary material to a macroscopic description in terms of constitutive parameters is not unique. Instead of following the textbook derivation of macroscopic Maxwell's equations, splitting electric and magnetic effects in distinct electric and magnetic constitutive parameters, one may include all the terms resulting from the averaging of the microscopic currents in a new definition of $\mathbf{D}$, dependent on $\mathbf{E}$ and its spatial derivatives (hence, nonlocal), while setting $\mathbf{B} = \mu_0 \mathbf{H}$ [98]. In this way, any magnetic and magneto-electric (i.e., chiral) effects can be regarded, equivalently, as (weak) forms of spatial dispersion of the permittivity. In broader terms, any bi-anisotropic linear material can be described by purely electric spatially dispersive constitutive relations [99]. One of the advantages of this equivalent representation is that Kramers-Kronig relations (a consequence of causality), and several related sum rules, apply to the function $\chi_e(\omega, \mathbf{k})$ for each individual wavevector. This allows deriving causality-based constraints and fundamental limits on the response of generic linear materials, as recently done in Ref. [100] to determine an upper bound on the refractive index of any transparent optical material and metamaterial.

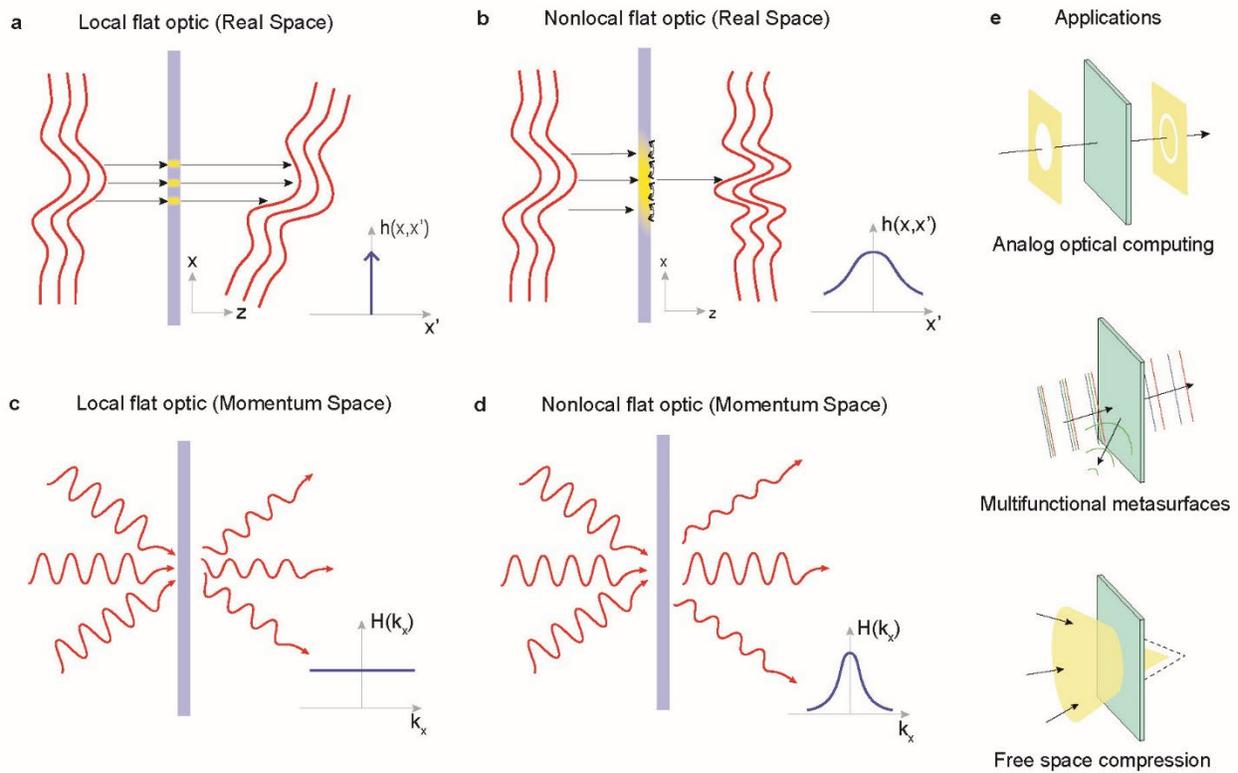

**Figure 1. Local vs. nonlocal flat optics.** (a,c) In a *perfectly* local metasurface, the optical response at a certain point on the output plane only depends on the applied field at a single point on the input plane, namely, the spatial impulse response $h$ is a delta function (a). This allows shaping the propagating wavefront as a function of transverse position locally and pointwise, which may result, for instance, in beam redirection. In momentum space (c), corresponding to a plane-wave expansion of the fields, the transfer function $H$ is therefore constant with respect to transverse wavenumber $k_x$ and, hence, angle of incidence. (These are idealizations, as the response is never perfectly local). (b,d) In a nonlocal metasurface, the optical response at each point on the output plane depends on the applied field in an extended spatial region. In momentum space, this corresponds to a non-constant transfer function, which allows shaping the propagating wavefront as a function of transverse momentum. (e) Examples of applications of nonlocal flat optics (see also Figs. 3,4). The new degrees of freedom afforded by strong effective nonlocality expand the design toolbox and enable functionalities that are impossible with individual local metasurfaces.

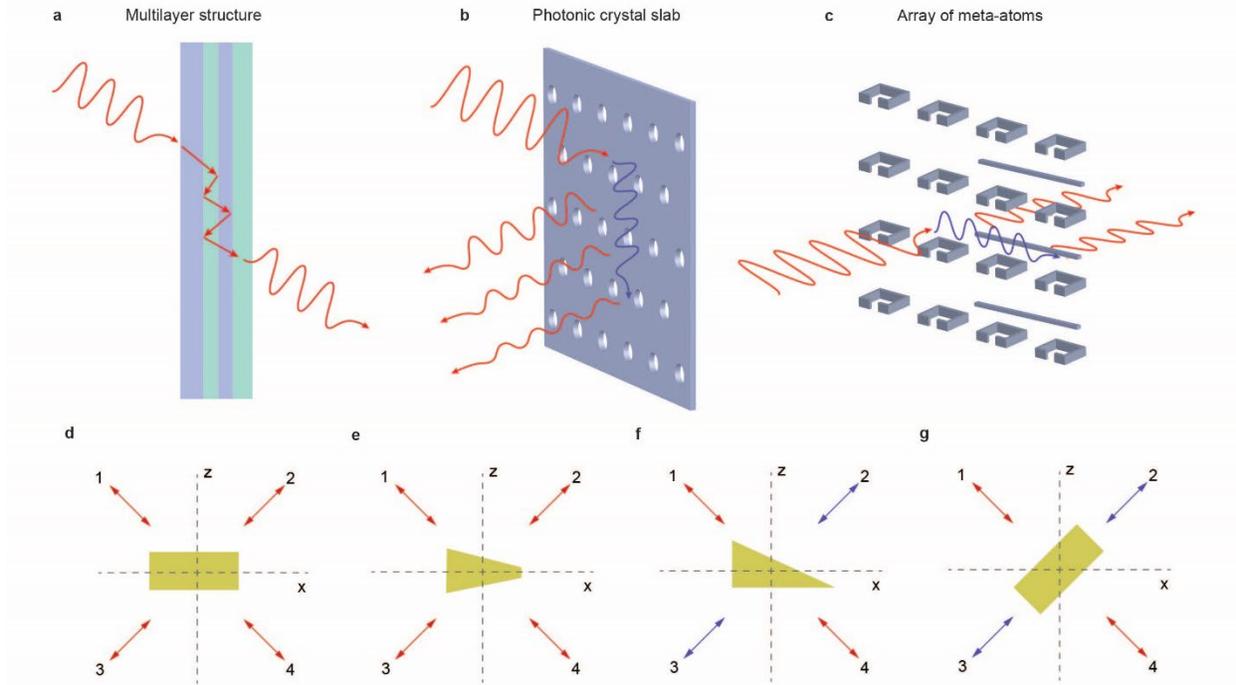

**Figure 2. Photonic platforms to realize strong artificial nonlocality.** (a,b,c) A strongly nonlocal response requires a wave of induced polarization traveling across the structure over relatively large distances to effectively "connect" the response of distant points. This can be achieved by relying on multiple reflections in multilayer structures (a) or via the excitation of a delocalized guided leaky mode, which naturally extends over an extended region of space, in gratings, photonic crystal slabs (b) or arrays of strongly interacting meta-atoms (c). In panels (b),(c), the guided leaky mode, denoted in blue, gradually radiates as it propagates, creating an indirect reflection/transmission pathway interfering with the direct one. These approaches may also be combined in the form of multilayer arrays or multilayer photonic crystal slabs, and may also include spatially varying structuring for local wavefront control. (d-g) Examples of meta-atoms in an array, as in (c), with different mirror symmetries. Double-headed arrows indicate four possible directions of incidence and transmission. If the structure is reciprocal, a different transmission response for opposite in-plane wavevectors requires breaking *both* horizontal- and vertical-mirror symmetries, as in the chiral meta-atom in (f), or the obliquely oriented anisotropic meta-atom in (g). Panels (c-g) are adapted from [17], APS.

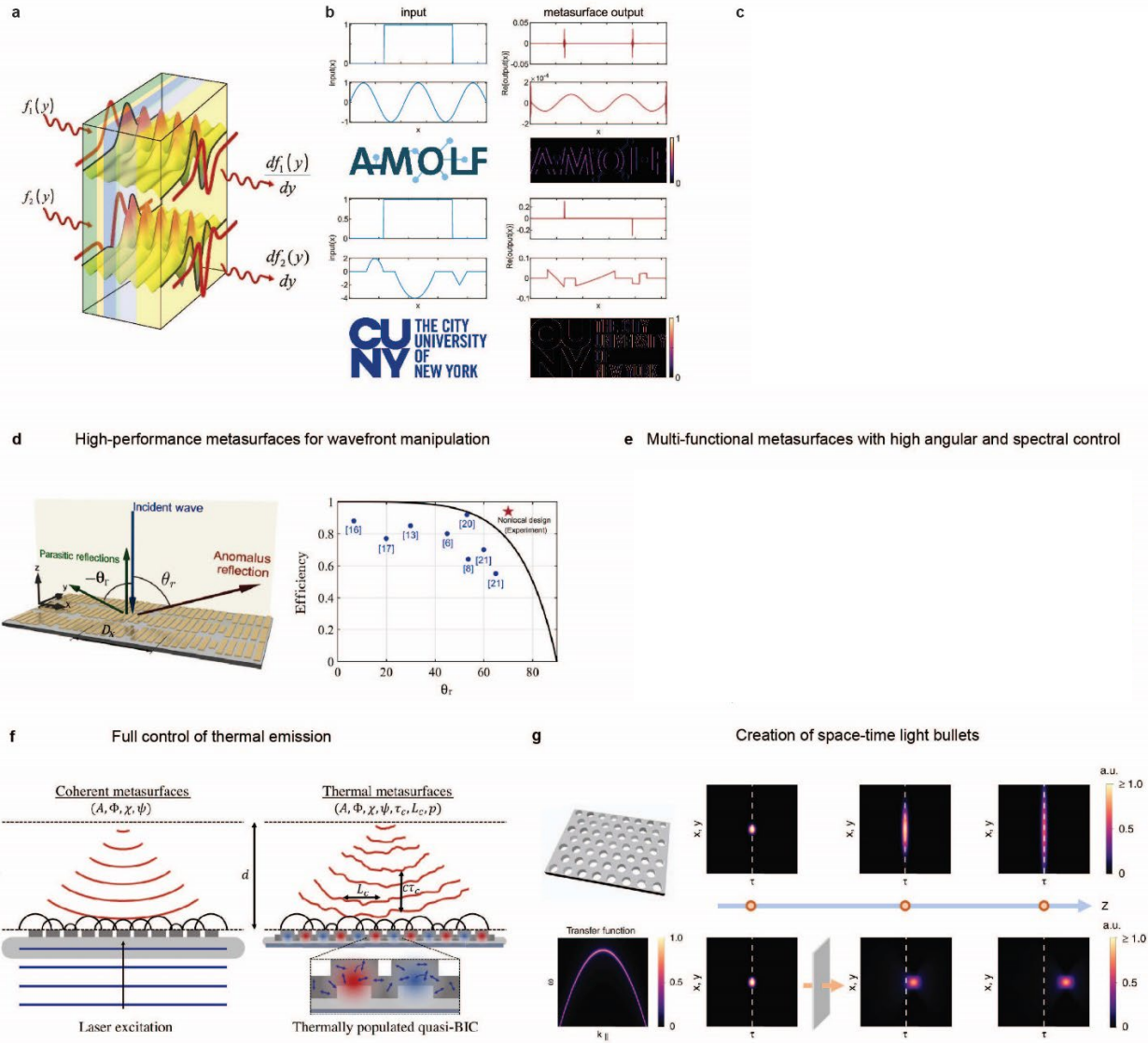

**Figure 3. Applications and opportunities enabled by nonlocal flat optics.** The new degrees of freedom enabled by nonlocality enable a plethora of new opportunities for light manipulation. (a,b,c) The wavevector-dependent response of nonlocal structures can be used for spatial-frequency filtering and to implement various spatial operators for analog optical computing, performing mathematical operations on a propagating wavefront (a), or analog image processing, e.g., edge detection (b,c). (d) The ability to channel energy transversely across a nonlocal metasurface via a guided leaky mode enables anomalous reflection/refraction and extreme wavefront transformation with much higher efficiency than possible with local metasurfaces. (e) The marked spatial and spectral dispersion of certain nonlocal structures allows manipulating light with high angular and frequency control. This feature can be used to create multifunctional metasurfaces producing spatially shaped wavefronts at different selected angles and wavelengths, with intriguing applications for augmented reality systems, e.g., glasses that provide an unperturbed view in the visible spectrum, while deflecting near-infrared light for eye-tracking. (f) Metasurfaces that combine nonlocal interactions, via a thermally populated guided mode, and local effects, determined by subwavelength structuring, can be engineered to provide full control over thermally emitted light. (g) The transfer function of a nonlocal structure, e.g., a suitable photonic-crystal slab, can be utilized to filter an incident pulse in frequency-wavevector domain and impart the required space-time correlation to create propagation-invariant optical wave packets known as light bullets. Adapted from: (a), ref. [25], AAAS; (b), ref. [33], ACS; (c), ref. [35], Springer Nature; (d), ref. [47], AAAS; (e), ref. [64], Springer Nature; (f), ref. [77], APS; (g), ref. [80], Springer Nature.

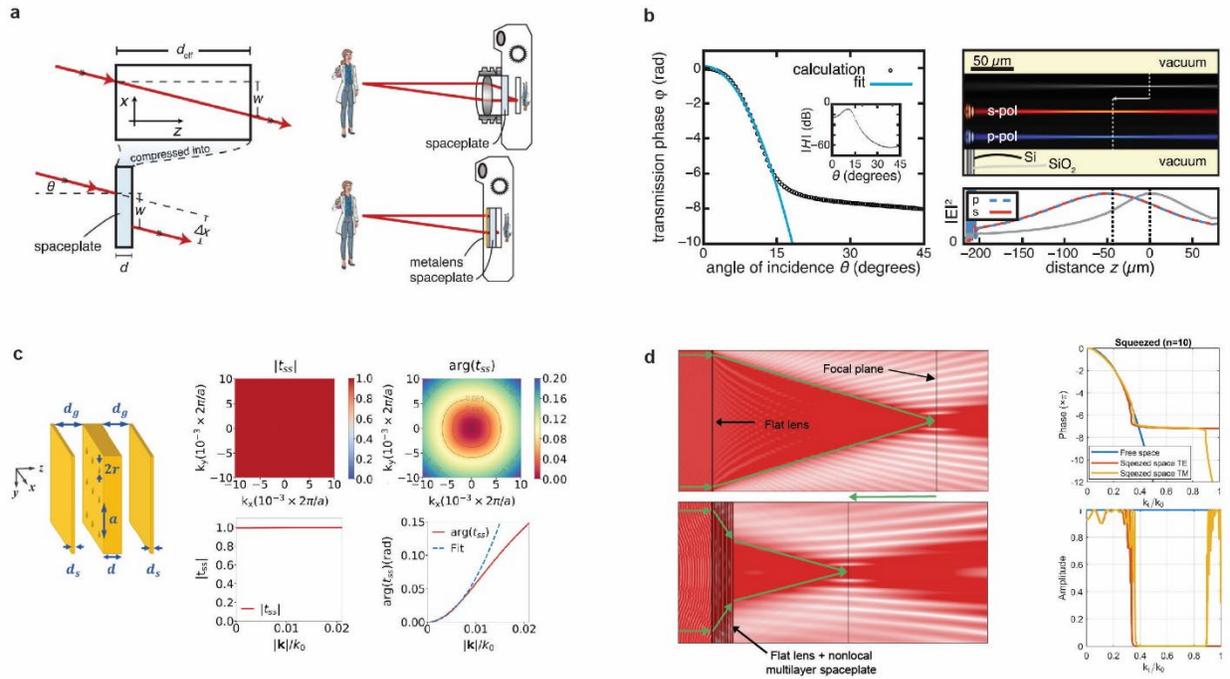

**Figure 4. Space compression with nonlocal spaceplates.** In addition to the applications highlighted in Fig. 3, nonlocal flat optics has also been recently used to create space-compression devices known as "spaceplates". (a) A spaceplate is a device that implements the angle-dependent optical response of a free-space volume over a smaller length, effectively compressing space for light propagation. The combination of metalenses and spaceplates may lead to the drastic miniaturization of any imaging system. (b,c,d) Spaceplates can be implemented using nonlocal structures such as photonic-crystal slabs (c) and multilayer thin films (b,d), engineered to have high transmission amplitude and an angle-dependent transmission phase that matches the response of a longer length of free space, over the largest possible angular and frequency ranges. As seen in the field-intensity maps in panel (d) a spaceplate causes the focus (and the entire field distribution) to move towards the (flat) lens, effectively decoupling the focal length of the lens from the actual distance at which focusing is achieved and, hence, the length of the imaging system. Adapted from: (a) and (b), ref. [26], Springer Nature; (c), ref. [37], OSA; (d), ref. [27], ACS.

Francesco Monticone and Kunal Shastri are at the School of Electrical and Computer Engineering, Cornell University, Ithaca, New York 14850, USA.

e-mail: francesco.monticone@cornell.edu

## Data availability

All relevant data are available from the corresponding author upon reasonable request.

## Competing interests

The authors declare no competing interests.

## Acknowledgements

The authors acknowledge support from the Air Force Office of Scientific Research with Grant No. FA9550-22-1-0204 through Dr. Arje Nachman.